\documentclass[a4paper,twoside]{article}
\usepackage[authoryear]{natbib}
\bibpunct{(}{)}{;}{a}{}{,}
\usepackage{graphicx}
\date{} 
\title{\large\bf\flushleft Optimal Multiwavelength Source Detection: Experience Gained from the WISE Mission}
\author{\parbox{\textwidth}{\flushleft
\vspace{-0.5cm}
{\it Kenneth A. Marsh and Thomas H. Jarrett}\\
\vspace{0.4cm}
{\small Infrared Processing and Analysis Center, California Institute of Technology 100-22, Pasadena, CA 91125}\\
{\small Email: kam@ipac.caltech.edu}}}
\begin{document}
\twocolumn[
\begin{changemargin}{.8cm}{.5cm}
\begin{minipage}{.9\textwidth}
\vspace{-1cm}
\maketitle
%
%
\small{\bf Abstract:}
We discuss the optimal detection of point sources from multiwavelength 
imaging data using an approach, referred to as MDET, which requires no prior 
knowledge of the
source spectrum.  MDET may be regarded as a somewhat more general
version of the so-called ``chi squared" technique.  We
describe the theoretical basis of the technique, and show examples of its
performance with four-channel infrared broad-band imaging data from the
WISE mission.  We also discuss the potential benefits of applying it to 
the multifrequency data cubes of the ASKAP surveys, and suggest that it
could increase the detection sensitivity of searches for neutral hydrogen
emission at moderately high redshifts.


\medskip
\medskip
\end{minipage}
\end{changemargin}
]
\small

\section{Introduction}

In many astronomical imaging applications, images are taken at multiple
wavelengths. Although the ability to detect faint sources can be
enhanced by stacking the images, a simple weighted linear combination 
produces a spectral bias dictated by the particular weighting function.
We describe a detection algorithm which overcomes this limitation, and
discuss its application to the {\em Wide-field Infrared Survey Explorer\/} 
(WISE) mission \citep{wright2010}.

Over a period of seven months from a precessing-polar orbit of the Earth,
WISE surveyed the whole sky in four infrared bands with
effective wavelengths of 3.4 $\mu$m (W1), 4.6 $\mu$m (W2), 12 $\mu$m
(W3) and 22 $\mu$m (W4) using an imaging array with a pixel
size of $2.75''$, and spatial resolution (FWHM) of approximately $6''$
at the three shortest wavelengths and $12''$ at the longest.
The WISE bands were chosen to optimize detection of
cool brown dwarfs and luminous infrared galaxies, but they are also well
placed to study most objects in the universe.  In general, the short
wavelength bands are sensitive to starlight, while
the long wavelength bands are sensitive to emission from the
interstellar medium and from dust associated with star formation 
(see, for example, \citet{jar11}).
We have performed source detection on the resulting
stacks of four-band images using the Multiband DETection (MDET) algorithm which
is optimal for the detection of
point sources in the presence of additive Gaussian noise.  In the context
of WISE, MDET represents the initial detection step in source photometry.
Its role is to produce a set of candidate sources which are then forwarded
to a separate module for detailed parameter estimation consisting of
source position, the flux at each band, the corresponding uncertainties,
and various measures of the estimation quality.

We discuss
our experience with MDET using WISE data, 
and discuss its potential benefits for source detection with the Australian 
Square Kilometre Array Pathfinder (ASKAP), whose design characteristics are
discussed by \cite{john09}.  In particular, we discuss its applicablilty
to the search for neutral hydrogen in various redshift ranges,
as planned for key projects\footnote{http://www.atnf.csiro.au/SKA/ssps.html} 
WALLABY (Widefield ASKAP L-Band Legacy
All-Sky Blind survey) and FLASH (The First Large Absorption Survey in HI).

\section{Theoretical Basis}

\subsection{Measurement Model}
The starting point for the detection step is the measurement model for an
isolated point source\footnote{In Subsection 2.3 we will discuss the
behavior in crowded fields, where this assumption is often violated}, 
assumed to be at location $\mathbf s$ and to have
flux $f_\lambda$ in the waveband denoted by index $\lambda$; it can be 
expressed as:
\begin{equation}
\rho_{\lambda i} = f_\lambda H_\lambda({\mathbf r}_{\lambda i} - {\mathbf s}) + 
b_{\lambda i} + \nu_{\lambda i}
\label{eq1}
\end{equation}
where $\rho_{\lambda i}$ is the observed value of the $i$th pixel at sky
location ${\mathbf r}_{\lambda i}$, $H_\lambda({\mathbf r})$ is the 
point spread function (PSF) representing the response of a focal-plane pixel
to a point source, $b_{\lambda i}$ is the background sky level, 
and $\nu_{\lambda i}$ is the noise, assumed 
to be a zero-mean Gaussian random process with covariance ${\mathbf C}_\nu$
defined by:
\begin{equation}
(C_\nu)_{\lambda i,\lambda' i'} \equiv
E\,\nu_{\lambda i} \nu_{\lambda' i'} = \delta_{\lambda \lambda'} 
\delta_{i i'}(\sigma_\nu)^2_{\lambda i} + (C_b)_{\lambda i, \lambda' i'}
\label{eq1a}
\end{equation}
where $E$ is the expectation operator, $(\sigma_\nu)_{\lambda i}$ is the 
standard deviation of measurement noise, assumed to be uncorrelated, and 
matrix ${\mathbf C}_{\rm b}$ represents the covariance of the background.

It is advantageous to estimate the background, $b_{\lambda i}$, ahead
of time (using, for example, median filtering with a window size, $W$, 
appropriate to the characteristic spatial scale of background variations) and
subtract its contribution, so that the measurement model may be rewritten:
\begin{equation}
\rho_{\lambda i} = f_\lambda H_\lambda({\mathbf r}_{\lambda i} - {\mathbf s}) + 
\nu_{\lambda i}
\label{eq2}
\end{equation}

If the observations are not sky background limited, or if the sky background
is flat, the fact that a background has been subtracted is not an issue.  
Otherwise, the analysis becomes complicated by the presence of spatial
correlations in the residuals after subtraction.  To minimize
the effects of such correlations, $W$
should be chosen comparable to the minimum spatial scale, $L$, of 
background fluctuations.  Provided the correlated component of background 
residuals is not too large in comparison to the measurement noise,
i.e., provided the diagnonal elements of ${\mathbf C}_b$ are smaller than
$(\sigma_\nu^2)_{\lambda i}$, then the residuals of sky background subtraction
can be lumped in with the measurement noise.  Specifically, the variance of
subtraction residuals can then be treated as a ``background confusion" term
which is added in quadrature to $(\sigma_\nu)^2_{\lambda i}$.  In the
case of WISE, this was a good approximation since the residuals of
background subtraction were dominated by 
low-level unresolved point sources and/or diffuse nebulosity on significantly
larger scales than the sources of interest.  The former component can be treated
as uncorrelated noise for present purposes, while the latter component
normally subtracts out cleanly.  An example involving background subtraction
in a region of heavy nebulosity is presented in Section 3.
In subsequent discussion we therefore assume that the residuals
of background subtraction can be treated in a similar fashion to
uncorrelated measurement noise, and denote 
the variance of combined noise as $\sigma^2_{\lambda i}$.

There could conceivably be situations in which
the above assumption would not be appropriate. For example, if background
fluctuations were of such a scale as to be difficult to distinguish from
genuine sources, then a more sophisticated approach would need to be taken.
This would not invalidate the detection approach to be described in the
next section, however, because the treatment could be extended to the
case of non-negligible background correlations in a straighforward way.
It would involve replacing the factor $1/\sigma^2_{\lambda i}$, which appears
in various summations, by $(C_\nu^{-1})_{\lambda i,\lambda' i'}$; the 
appropriate summations would then be performed over the set of 
$\lambda, i, \lambda', i'$ rather than simply $\lambda, i$.

\subsection{Detection Algorithm}
Based on the measurement model expressed by Equation (\ref{eq2}), the source 
detection procedure 
involves comparing the relative probabilities of the following two
hypotheses at each location, ${\mathbf s}$, within a predefined regular grid of
points on the sky:

Hypothesis (A):  ${\mathbf s}$ lies on blank sky at all wavelengths

Hypothesis (B):  ${\mathbf s}$ represents the location of a source whose 
flux densities are the most probable values, denoted by $\hat{f_\lambda}$. 

\subsubsection{Prior information on flux values: the positivity
constraint}

To compare the above hypotheses requires knowledge of
$\hat{f_\lambda}$, which we obtain by maximizing the conditional probability,
$P({\mathbf f}|\rho)$, with respect to ${\mathbf f}$, where ${\mathbf f}$ is a 
vector whose components are the set of $f_\lambda$, and $\rho$ is a vector 
whose components are the set of 
pixel values, $\rho_{\lambda i}$, in the vicinity of ${\mathbf s}$.
The conditional probability itself is given by Bayes' rule, i.e.,
\begin{equation}
P({\mathbf f}|\rho) = P(\rho|{\mathbf f})P({\mathbf f})/P(\rho)
\label{eq3a}
\end{equation}
where
\begin{equation}
\ln P(\rho|{\mathbf f}) = -\frac{1}{2}\sum_{\lambda ,i} 
\frac{1}{\sigma_{\lambda i}^2} \left[\rho_{\lambda i} - 
f_\lambda H_\lambda({\mathbf r}_{\lambda i} - {\mathbf s})\right]^2
\label{eq3b}
\end{equation}
and $P({\mathbf f})$ represents our {\em a priori\/} knowledge about possible
flux values.  We make no {\em a priori\/} assumptions about relative 
probabilities of spectral shapes of astrophysical objects, so our 
$P({\mathbf f})$ is completely neutral on that point, i.e., we do not
wish to introduce any color biases into the detector.  However, one
important piece of knowledge that we {\em do\/} have is that 
flux is positive.  We can thus express $P({\mathbf f})$ as:
\begin{equation}
P({\mathbf f}) = \left\{ \begin{array}{ll}
	{\rm const.} & \mbox{if $f_\lambda \ge 0 \quad\forall \lambda$} \\
	0            & \mbox{otherwise}
			\end{array}
		\right.
\label{eq3c}
\end{equation}

The remaining quantity, $P({\rho})$, in (\ref{eq3a}), represents a 
normalization factor.  The maximization of $P({\mathbf f}|\rho)$ then yields:
\begin{equation}
\hat{f_\lambda} = \theta\left([\,\sum_i \frac{1}{\sigma_{\lambda i}^2}
H_{\lambda i}\rho_{\lambda i}\,]/ \sum_i \frac{1}{\sigma_{\lambda i}^2}
H_{\lambda i}^2 \right)
\label{eq3d}
\end{equation}
where the point spread function has been abbreviated to $H_{\lambda i}\equiv H_\lambda({\mathbf r}_{\lambda i} - {\mathbf s})$,
and $\theta(x)$ represents a function which is equal to its argument if the
latter is nonnegative and 0 otherwise.  The summations in (\ref{eq3d}) are
over all pixels within a predefined neighborhood of ${\mathbf s}$.  

\subsubsection{Evaluating the relative probabilities}

With a further application of Bayes' rule, we can now express the 
probabilities of hypotheses (A) and (B), above, as:
\begin{equation}
P({\rm sky}|\rho,m_0) = P({\rm sky},m_0) \prod_{\lambda} P(\rho_\lambda|{\rm sky},m_0)/P(\rho_\lambda,m_0) 
\label{eq4}
\end{equation}

\begin{equation}
P(\hat{\mathbf f}|\rho,m) = P(\hat{\mathbf f},m)
\prod_{\lambda} P(\rho_\lambda|\hat{\mathbf f},m)/P(\rho_\lambda,m)
\label{eq5}
\end{equation}
where $m_0$ represents the sky-only model corresponding to hypothesis (A),
and $m$ represents the model corresponding to hypothesis (B), based on
Equation (\ref{eq2}). 

The likelihoods $P(\rho|{\rm sky},m_0)$ and $P(\rho|\hat{f_\lambda},m)$
are given by:
\begin{equation}
\ln P(\rho_\lambda|{\rm sky},m_0)=-\frac{1}{2}\sum_i\frac{\rho_{\lambda i}^2}
{\sigma_{\lambda i}^2} + {\rm const.}
\label{eq6} 
\end{equation}

\begin{equation}
\ln P(\rho_\lambda|\hat{f_\lambda},m) = -\frac{1}{2}\sum_i
\frac{1}{\sigma_{\lambda i}^2}(\rho_{\lambda i} - 
\hat{f_\lambda}H_{\lambda i})^2 + {\rm const.}
\label{eq7}
\end{equation}

Assuming that we have no prior knowledge about the possible presence or
absence of a source at ${\mathbf s}$,
all of the other factors in (\ref{eq4}) and (\ref{eq5}) may be regarded
as constants for present purposes,
and we can thus express the probability ratio (source/sky) as:
\begin{equation}
\ln\,\frac{P(\hat{\mathbf f}|\rho,m)}{P({\rm sky}|\rho,m_0)} =
\frac{1}{2}\sum_\lambda\hat{f_\lambda}^2\sum_i\frac{H_{\lambda i}^2}
{\sigma_{\lambda i}^2} + {\rm const.}
\label{eq8}
\end{equation}

Substituting for $\hat{f_\lambda}$ using (\ref{eq3d}), we can express this
probability ratio as:
\begin{equation}
\ln\,\frac{P(\hat{\mathbf f}|\rho,m)}{P({\rm sky}|\rho,m_0)} =
\frac{1}{2}\phi({\mathbf s})^2 + {\rm const.}
\label{eq9}
\end{equation}
where $\phi({\mathbf s})$ is defined as:
\begin{equation}
\phi({\mathbf s}) = \left(\sum_\lambda\frac{\theta\!\left(\,\sum_i
(\rho_{\lambda i}/\sigma_{\lambda i}^2) 
H_\lambda({\mathbf r}_{\lambda i}\!-\!{\mathbf s})\,\,\right)^2}
{\sum_i (1/\sigma_{\lambda i}^2) 
H_\lambda({\mathbf r}_{\lambda i}\!-\!{\mathbf s})^2}\right)^\frac{1}{2}
\label{eq12}
\end{equation}
in which we have replaced the values of the point spread function, 
$H_{\lambda i}$, by their more explicit form. 

\subsubsection{Criterion for source detection}
From (\ref{eq9}), maxima in $\phi({\mathbf s})$ correspond to maxima in
the (source/sky) probability ratio, and hence an image formed by calculating
$\phi({\mathbf s})$ over a regular grid of positions, ${\mathbf s}$, 
would be a suitable basis for optimal source detection.  It is apparent
from (\ref{eq12}) that such an image represents a quadrature sum of
matched filters at the individual wavelengths, with appropriate normalization.
The noise properties of such an image can be assessed by expressing
$\phi({\mathbf s})$ in terms of the 
{\em a posteriori\/} variance of $\hat{f_\lambda}$, given by:
\begin{equation}
(\sigma_f^2)_\lambda = 1/\sum_i\frac{H_{\lambda i}^2}{\sigma_{\lambda i}^2}
\label{eq10}
\end{equation}
from which we obtain:
\begin{equation}
\phi({\mathbf s}) = \left(\sum_\lambda
\frac{\hat{f_\lambda}^2}{(\sigma_f^2)_\lambda}\right)^\frac{1}{2}
\label{eq11}
\end{equation}

It is readily shown, from (\ref{eq11}), that the standard deviation of 
$\phi({\mathbf s})$ is unity, i.e., $\phi({\mathbf s})$ itself is in units of 
standard deviations.
Therefore, for a given detection threshold $T_{\rm d}$ [sigmas], the most
likely locations of sources correspond to those for which
$\phi({\mathbf s})\ge T_{\rm d}$. The quantity $T_{\rm d}$ represents our 
pre-defined detection threshold, which for the WISE Preliminary 
Release \citep{cutri2011} was set at 7, i.e., all peaks above a signal 
to noise ratio of 7 were taken as candidate sources which
were then passed to the photometry module for precise estimation
of flux and position.

\subsubsection{Summary of the detection procedure}

The multiwavelength detection algorithm as derived above consists of the
following steps:
\begin{itemize}
  \item[1.] Subtract a slowly varying background from the images at
each of the individual wavelengths in order to ``flatten" the sky.
  \item[2.] Calculate a spatial matched filter image at each individual
wavelength.  The result, obtained by cross-correlating the observed image 
with the PSF, optimizes the $S/N$ of the point sources in the image at
that wavelength.
  \item[3.]  Divide each such matched filter image by a corresponding 
uncertainty image representing the spatial variation in the standard deviation 
of local background noise.
  \item[4.]  Set negative pixel values in the resulting image to zero.
  \item[5.]  Form the quadrature sum of the clipped images.
  \item[6.]  Threshold this image at the desired signal-to-noise level,
$T_{\rm d}$, and find all local maxima in the thresholded image.
\end{itemize}

\begin{figure}[h]
\begin{center}
\includegraphics[scale=0.45, angle=0]{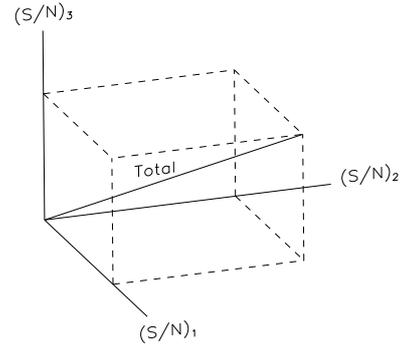}
\caption{Graphical illustration of the effect of combining matched filter 
images at multiple wavelengths, three in this case.  The 
axes in the figure represent the pixel values corresponding to the source
peak in the three individual bands, in units of the standard deviation of
local measurement noise.  When the corresponding matched filter images
are combined in quadrature, as per the MDET procedure, that pixel
receives a value indicated by the length of the vector ``Total."
Since the signals in the individual bands 
are orthogonal, noise in individual bands has minimal effect on the resultant, 
and the latter is independent of which particular bands have the
largest $S/N$, i.e., there is no bias towards any particular spectral shape.}
\label{fig1}
\end{center}
\end{figure}

A graphical illustration of the image combination procedure is given
in Figure \ref{fig1}.  

The MDET detection algorithm is similar to one proposed by \citet{szal1999},
often referred to as the ``chi squared" method. The central operation,
in both cases, is a quadrature sum of matched filter images.  
An important difference between the two procedures, however, is the
fact that in our technique we threshold the matched filter images at zero before
squaring and combining, thus avoiding the contaminating effect of
squared negative values in the image sum.  This is a direct consequence
of our prior information concerning the positivity of intensity, imposed
via our Bayesian framework using Equation (\ref{eq3c}). It reduces the
background noise on the combined image by a factor of $\sqrt{2}$, and 
therefore increases the sensitivity by the same factor. In applying the
positivity constraint, however, care must be exercised in the background 
subtraction step, particularly in a confused region in which the background 
may vary on relatively short spatial scales.

\subsection{Allowance for confusion}
In the above derivation, the assumption was made that the images
of adjacent sources do not overlap.  Although all matched
filters are subject to this limitation, the effects of confusion can
be can be more acute in the multiwavelength case.  For example, 
in a WISE combined detection image, a star in W1 may become confused 
with an extended source such as a galaxy in W4.  We overcome such effects 
by supplementing the set of multiband detections with the 
results of single-band detections, thereby maintaining the 
increased sensitivity of multiband detection while not missing any 
detections due to cross-band blending effects.  Most WISE sources, however,
are detected in the multiband step; the
number of supplementary single-band detections is typically only $\sim1$\%
of the total, and many of those are simply the result of noise bumps.  We 
include them for considerations of completeness.

The fact that the mathematical formalism on which MDET is based makes no
explicit allowance for confusion is not a problem within the overall
scheme of WISE source photometry, since MDET is simply the initial stage of
a procedure in which the results are subsequently refined in a parameter 
estimation step.
For example, if two closely spaced components are blended into a single
peak in the MDET detection image, they are subsequently separated by
the so-called ``active deblending" procedure in profile-fitting photometry
\citep{cutri2011}.  Such a situation is signaled by the presence of
an elevated value of the reduced chi squared, $\chi_\nu^2$, of the 
maximum likelihood fit and indicates that an extra point-source component 
must be added to the model.
Extended sources represent another violation of the
assumptions of the detection algorithm, although this was not a serious issue 
for WISE since most of the sources were spatially unresolved. Modification
of the algorithm to optimize MDET for the detection of faint extended
sources would require the use of a set of extended source templates
in Equation (\ref{eq12}) instead of the point spread function,
$H_\lambda({\mathbf r})$.

\section{Examples of Application to WISE Data}

\begin{figure}[h]
\begin{center}
\includegraphics[scale=1.0, angle=0]{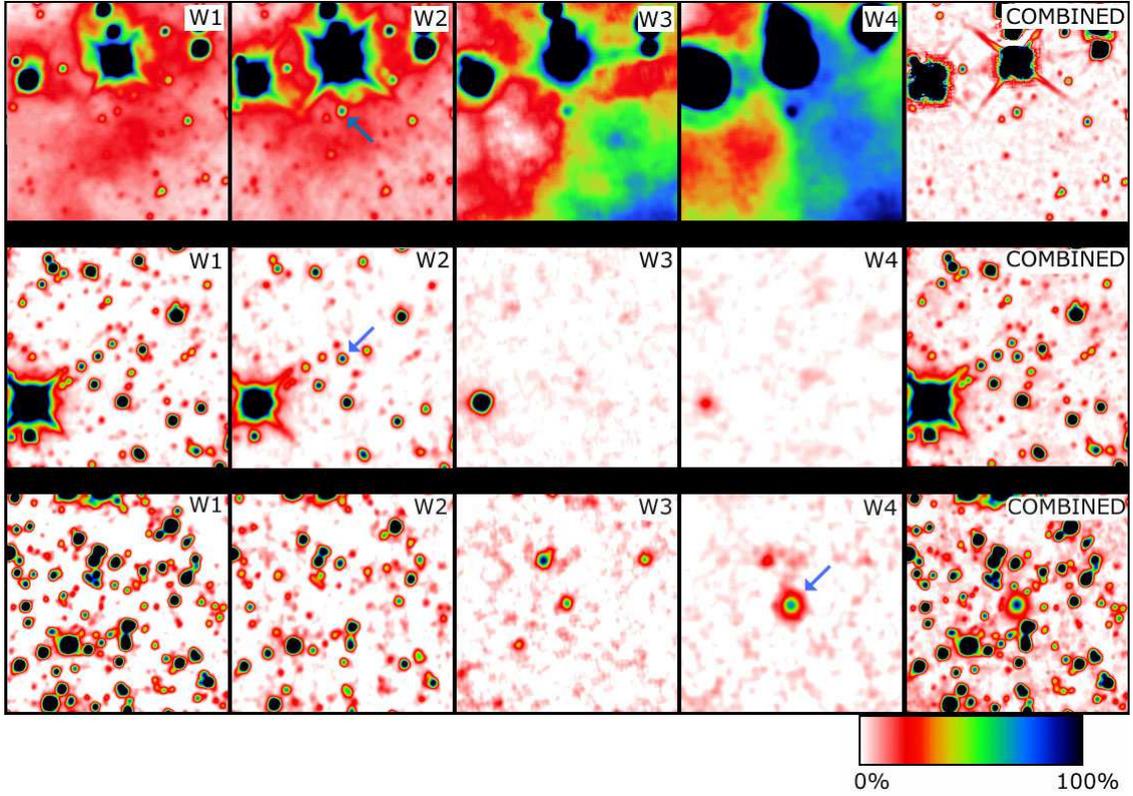}
\caption{Examples of the use of MDET for combining images at multiple
wavelengths to increase detection sensitivity.
{\em Top row:\/} Young stellar object (YSO); {\em Middle row:\/} Ultra-cool 
brown dwarf (BD);  {\em Bottom row:\/} Hyper-Luminous InfraRed Galaxy (HyLIRG). 
Coadded images, each with $5'\times5'$ field of view, are shown for 
the four WISE wavebands, W1 (3.4 $\mu$m), W2 (4.6 $\mu$m), W3 (12 $\mu$m),
and W4 (22 $\mu$m); these were combined to produce the optimal detection
image (labeled ``COMBINED") shown at the right of each row. For each
of the three objects, all five images are presented on the same (linear)
intensity scale in units of $S/N$, whose peak values are 
102, 68, and 33 for the YSO, BD, and LyLIRG, respectively. In each case,
the object of interest is indicated by the blue arrow for the 
waveband with highest $S/N$ (W2 for the YSO and brown dwarf; W4 for the 
HyLIRG).}
\label{fig2}
\end{center}
\end{figure}

Figure \ref{fig2} shows three examples in which WISE images at four wavelength
bands have been combined to produce a detection image using the MDET procedure
described above. These examples serve to illustrate several aspects of
the procedure.

The first example is of a young stellar object (a class I protostar candidate)
in the L1689 starforming cloud of $\rho$ Oph.  It is visible in all four 
WISE bands against a background of diffuse emission.

The second example is of an ultra-cool brown dwarf, using data from 
\citet{mainzer2011}.
This object (WISEPC J04583.90+643451.9) has an estimated temperature of 600 K
and a spectral class T9.  At temperatures such as these, the spectrum is
sharply peaked near 4.5 $\mu$m, corresponding to a relatively 
narrow ``island" of low opacity between the 
heavy absorption due to such molecular components as water and methane.
For this reason, the source is by far the most prominent in the
W2 waveband of WISE.  The W1 and W2 filters were, in fact, optimized for
the detection of this feature.

The third example is of a Hyper-Luminous InfraRed Galaxy (HyLIRG), using
data from Eisenhardt et al. (2011, in preparation).  This is a very red 
object, and is thus brightest in W4.

In all three cases, the images in the individual bands have 
been convolved with the respective PSFs to produce, in essence, optimal 
matched filter images at those wavelengths, so the corresponding $S/N$ values 
are directly comparable with that of
the combined (detection) image.  The actual $S/N$ values are presented
in Table \ref{tbl-1}.  

\tabcolsep 0.5pt
\begin{table}[h]
\begin{center}
\caption{Detection $S/N$ for WISE observations of three selected objects} \label{tbl-1}
\begin{tabular}{lccccc}
\hline {\small Object} & {\small W1} & {\small W2} & {\small W3} & {\small W4} & {\small Combined}\\
 & {\small (3.4 $\mu$m)} & {\small (4.6 $\mu$m)} & {\small (12 $\mu$m)} & {\small (22 $\mu$m)} \\
\hline 
{\small YSO} & {\small 23.67} & {\small 65.07} & {\small 33.27} & {\small 25.89} & {\small 101.89} \\
{\small BD} & {\small 9.33} & {\small 67.48} & {\small 2.38} & {\small 0.03} & {\small 68.16} \\
{\small HyLIRG} & {\small 2.11} & {\small 1.12} &  {\small 22.69} & {\small 24.35} & {\small 33.15} \\
\hline
\end{tabular}
\medskip\\
\end{center}
\end{table}

The first example (YSO) demonstrates the
improvement in detectability that results from combining the images at
all four wavelengths.  As discussed above, one of the steps involved in
this procedure is to subtract a slowly-varying sky background.
The latter was estimated by median filtering using a moving window of
size $21'' \times 21''$, chosen to be representative of the spatial scale
of the background variations.
Note that the combined $S/N$ (101.9) exceeds
the quadrature combination of the $S/N$ at the individual bands (81.1);
the additional improvement is due to the effect of the positivity constraint.

Such a gain in $S/N$ is not obtained in the second example (BD)
since the source is detected primarily 
\clearpage

\clearpage

\noindent in one band only.  The example does,
however, demonstrate that non-detection in the three ``dropout" bands did
not contaminate the detection.  The third example (HyLIRG)
represents a combination of both circumstances, since the source is detected in
only two of the four bands.  Nevertheless, the two bands in which the
source is detected have served to increase the source $S/N$ in the combined
image, while the two dropout bands have not appreciably contaminated
the result.

Overall, the use of MDET as the initial detection step has benefited
WISE photometry in the following ways:
\begin{itemize}
\item[1.] The detection sensitivity has been increased for the reasons
discussed above.  This increase does not come at the expense of reliability,
since the image combination process tends to average
out the noise bumps.  The inferred reliability meets the
WISE Functional Requirement value of 99.9\% for SNR $>$ 20 \citep{cutri2011}.
\item[2.] The fact that a single position is supplied for each source, even
though the source is detected in multiple bands, facilitates simultaneous
multiband parameter estimation.  The advantages of the latter are that
no bandmerging is required (thus avoiding band-to-band matching ambiguities 
crowded fields), and that dropout bands automatically receive a
flux upper limit.
\end{itemize}

\section{Application to ASKAP}

MDET can facilitate the optimal detection of faint sources by combining
images along the frequency axis of an ASKAP data cube without introducing 
biases which favor one spectral shape over another. However, because of the 
limited frequency range of ASKAP (700-1800 MHz receiver range; 300 MHz 
correlator bandwidth), the benefits of the algorithm
would be realized to a much larger extent for sources whose spectra contain
narrow-band features than for sources with broadband spectra.
As an example of the latter, consider a typical AGN whose flux density spectrum
in the ASKAP frequency range is of the approximate form $S_\nu\propto\nu^{-1}$.
Although the combining of images over all ASKAP channels will increase the
$S/N$ by approximately the square root of the number of channels involved,
the same will also be true for a uniformly-weighted linear combination of
images of such a source.  In fact, for a source with a similar spectral
index, the improvement in $S/N$ afforded by MDET over that obtained with
the linear combination would be less than 4\%.  
The situation is radically different for spectroscopy, however, and we
now discuss an important potential application.

A major scientific goal for ASKAP and SKA is the detection of neutral
hydrogen in distant galaxies, an important aspect of the study of galaxy
formation and evolution \citep{john08,raw04}. The MDET procedure could
be used to advantage in such searches by providing 
a spectrally neutral way to combine the images for
all narrow-band channels within the frequency range corresponding to
a given range of redshifts.  As an example, we consider the
combination of signals from a WALLABY data cube whose spectral  
sampling interval corresponds to 4 km s$^{-1}$.  We suppose that somewhere in
the redshift range $z<0.1$ is the signal from a velocity-broadened HI line, 
a typical width for which might be $\sim400$ km s$^{-1}$ \citep{kori96}.
Since the achievable gain in detectability
increases monotonically with the per-channel $S/N$ of the
spectral peaks of the source, it is desirable to carry out the
detection step at a spectral resolution commensurate with the spectral
structure of interest, so that some boxcar averaging of
the spectral channels may be warranted. In the present example, combining 
the raw channels in groups of 25 would result in an effective
channel width of 100 km s$^{-1}$, so that our source signal would
be spread over 4 such channels.  For a peak $S/N$ per channel $\sim10$, an 
unweighted linear combination of these channels would then produce a signal 
with $S/N\sim2$, i.e., barely detectable, whereas the MDET procedure would 
result in  $S/N\sim12$.  Of course, the {\em peak\/} signal from a 
properly-tuned spectral matched filter\footnote{As distinct from the 
{\em spatial\/} matched filter which is an integral part of the MDET 
procedure} would be even greater, but we would then be optimizing for a 
particular line width and shape, and be less sensitive to other 
potentially interesting structure. 
For example, if optimized for 100 km s$^{-1}$ structure, the spectral matched
filter could do no better than  $S/N\sim10$.  While this numerical
example was illustrative only, we can make the general statement that
by combining images from WALLABY data cubes using the MDET procedure, 
we obtain an optimal answer to the question: In which portions of the field 
of view are the observations inconsistent with random noise?

With an appropriate sign change in Equation (\ref{eq12}), the procedure
could be applied to the detection of HI absorption, and therefore be of
potential benefit to the processing of data from the FLASH project in the 
redshift range $0.5<z<1.0$.
Ultimately, it is expected that the full SKA will enable detection out to 
$z\sim2.5$ \citep{zwaan06}.  At such high redshifts, the emission is too weak 
to detect the HI clouds of individual galaxies.  The problem can be mitigated,
however, by combining the HI images of overlapping clouds of galaxies with known
redshifts \citep{khandai2011}.  This approach enabled \citet{chang2010} to
detect neutral hydrogen out to $z=0.8$.  MDET has the potential for further
improving this technique, since it provides a way of combining the
images without prior knowledge of the redshifts of individual clouds.

\section{Conclusions}

MDET provides a procedure for combining the narrow-band images 
within a data cube in such a way as to increase optimally the detection
signal to noise ratio without introducing a color bias.  We have used
it successfully in the WISE mission, and believe it will be of substantial
benefit to ASKAP, particularly for sources with narrow-band spectral features.
In particular, we suggest that it could aid in the detection of neutral 
hydrogen in distant galaxies.

\section*{Acknowledgments} 

We thank Dr. B. Koribalski for helpful comments. 
This publication makes use of data products from the Wide-field Infrared
Survey Explorer, which is a joint project of the University of
California, Los Angeles, and the Jet Propulsion Laboratory/California
Institute of Technology, funded by the National Aeronautics and Space
Administration.


\begin{thebibliography}{}
\bibitem[Chang et al.(2010)]{chang2010} Chang, T.-C., Pen, U.-L., Bandura, K.,
    \& Peterson, J. B. 2010, Nature, 466, 463
\bibitem[Cutri et al.(2011)]{cutri2011} Cutri, R. M. et al. 2011,
    ``Explanatory Supplement to the WISE Preliminary Release,"
     http://wise2.ipac.caltech.edu/docs/release/prelim/expsup
\bibitem[Jarrett et al.(2011)]{jar11} Jarrett, T. H. et al., 2011, ApJ,
    735, 112
\bibitem[Johnston et al.(2008)]{john08} Johnston, S. et al. 2008, ``Science
    With ASKAP," arXiv:0810.5187
\bibitem[Johnston et al.(2009)]{john09} Johnston, S., Feain, I. J.,
    \& Gupta, N. 2009, in ``The Low-Frequency Radio Universe," ASP Conf.
    Series, Vol. LFRU, 2009; Eds: D.J. Siakia, Dave Green, Y Gupta and
    Tiziana Venturi  
\bibitem[Khandai et al.(2011)]{khandai2011} Khandai, N., Sethi, S. K., 
    DiMatteo, T. \& 4 coauthors, 2011, MNRAS, in press
\bibitem[Koribalski(1996)]{kori96} Koribalski, B. 1996, ASPC, 106, 238
\bibitem[Mainzer et al.(2011)]{mainzer2011} Mainzer, A., Cushing, M. C.,
    Skrutskie, M. \& 16 co-authors, 2011, ApJ, 726, 30
\bibitem[Rawlings et al.(2004)]{raw04} Rawlings, S., Abdalin, F. B., Bridle,
    S. L. \& 4 coauthors, 2004, in ``Science with the Square Kilometre Array"
\bibitem[Szalay et al.(1999)]{szal1999} Szalay, A. S., Connolly, A. J., \&
    Szololy, G. P. 1999, AJ, 117, 68
\bibitem[Wright et al.(2010)]{wright2010} Wright, E. L. et al. 2010, AJ,
    140,1868
\bibitem[Zwaan(2006)]{zwaan06} Zwaan, M. 2006, in ``Cosmology, Galaxy
    Formation and Astroparticle Physics on the Pathway to the SKA,"
    Kl\"{o}ckner, H.-R., Jarvis, M. \& Rawlings, S. (eds), Oxford, U.K.
\end{thebibliography}
\end{document}